\documentclass[prl,preprint,amsmath,amssymb,superscriptaddress]{revtex4}
\usepackage{graphicx}
\usepackage{natbib}
\usepackage{latexsym}
\usepackage{hyperref}

\begin{document}

\title{Inhibited, Explosive and Anisotropic Relaxation in a Gas of Molecular Super-Rotors}

%\maketitle
%\today

\author{Yuri Khodorkovsky}
\affiliation{Department of Chemical Physics, The Weizmann Institute of Science,
Rehovot 76100, Israel}

\author{Uri Steinitz}
\affiliation{Department of Chemical Physics, The Weizmann Institute of Science,
Rehovot 76100, Israel}

\author{Jean-Michel Hartmann}
\affiliation{Laboratoire Interuniversitaire des Syst\`{e}mes Atmosph\'{e}riques (LISA) CNRS (UMR 7583), Universit\'{e} Paris Est Cr\'{e}teil, Universit\'{e} Paris Diderot, Institut Pierre-Simon Laplace, Universit\'{e} Paris Est Cr\'{e}teil, 94010 Cr\'{e}teil Cedex, France}

\author{Ilya Sh.\ Averbukh \footnote{corresponding author: ilya.averbukh@weizmann.ac.il}}
\affiliation{Department of Chemical Physics, The Weizmann Institute of Science,
Rehovot 76100, Israel}

\date{\today}
\maketitle
%\today

\newpage
\textbf{Recently, several femtosecond laser techniques have been developed~ \cite{centrifuge_corkum1,centrifuge_corkum2,Mullin1,Mullin2,Valery,Bucksbaum,NJP,Kitano_udr,Khodorkovsky_udr_benzene,Floss_prl}
that are capable of bringing gas molecules to extremely fast rotation in a very short time, while keeping their translational motion intact and relatively slow.
We investigate collisional equilibration dynamics of this new  state of  molecular gases, and find that it follows a remarkable generic scenario. The route to equilibrium starts with a durable metastable `gyroscopic stage', in the course of which the molecules maintain their fast rotation and orientation of the angular momentum through many collisions. The inhibited rotational-translational relaxation is characterized by a persistent anisotropy in the molecular angular distribution, and  is  manifested in the long-lasting optical birefringence, and anisotropic diffusion in the gas.  After a certain induction time, the `gyroscopic stage' is abruptly terminated by a self-accelerating explosive rotational-translational energy exchange leading the gas towards the final thermal equilibrium. We illustrate our conclusions by direct Molecular Dynamics (MD) simulation of super-rotors in several gases consisting of common linear molecules (such $N_{2}$, $O_{2}$ and $CO_{2}$).
}

The efficiency of the collisional energy transfer from the internal
molecular degrees of freedom (vibrational or rotational) to the translational
ones is crucially dependent on the typical time scales of the processes
involved. To analyze the rotational-translational (RT) energy transfer
between colliding molecules in a gas, one has to compare the typical
rotational period, $t_{rot}=2\pi/\omega_{rot}$ ($\omega_{rot}$ is
the angular rotational frequency) and the characteristic duration
of a collision, $t_{coll}$. The latter may be estimated as the time
it takes for a molecule of the typical speed $v$  to traverse the range $\sigma$
of the intermolecular potential:  $t_{coll}\sim2\sigma/v$.  The
ratio of these two time scales is the so-called adiabaticity parameter
$a=t_{coll}/t_{rot}$. When the $t_{coll}$ is long enough for the
molecule to complete several revolutions during the collision, i.e.
$a\gg1$, the collision is adiabatic with respect to the rotational
dynamics of the molecule. During the encounter of two molecules, the
molecular rotational dynamics adiabatically adjusts itself to a
slowly varying intermolecular interaction, and the rotational energy
of the molecules returns to its initial value after the end of the
collision. In such cases the RT relaxation is suppressed, and the
fast rotating molecules are dubbed super-rotors (SR)\cite{Stwalley2000}.
On the other hand, when $a\sim 1$ (or smaller), the collisions are impulsive, and
efficient energy transfer occurs between the rotational and the translational
degrees of freedom. Similar arguments are relevant also to the vibrational-translational
relaxation process, and they are known since the introduction of the
Landau-Teller theory \cite{LandauTeller} describing the vibrational-translational
energy exchange in a collision between an atom and a diatomic molecule.
The main difference between rotations and vibrations is the much smaller
typical timescale of the latter, which makes the 'adiabatic regime' more accessible
for the vibrations. Additionally, the dependence of rotational
frequency on energy provides a threshold for adiabaticity. It has
been known for a while, that highly rotationally excited molecules
appearing in some exothermic chemical reactions and photodissociation
processes may exhibit extended collisional relaxation \cite{Polanyi}.
More recently, it has been predicted that trapped ultra-cold molecules
may also experience adiabatic suppression of the rotational relaxation
\cite{Stwalley2000,Forrey2001} which may open new options for cold chemistry (see~\cite{Forrey2011} and references therein).
 In the latter case the translational
temperatures are very low, and the adiabaticity is enhanced by the
prolonged  duration of the collisions that are quantum in nature.
Nowadays,  molecules
can be excited to become SRs by optical techniques aiming at inducing
molecular alignment by short laser pulses (for reviews see Refs.~\cite{Stapelfeldt_review,Ohshima,my_review}),
especially utilizing the schemes using multiple laser pulses~\cite{Bucksbaum}.
Moreover, several methods have been suggested and demonstrated for
converting the transient molecular alignment into a concerted unidirectional
molecular rotation, including the techniques of ``optical centrifuge''~\cite{centrifuge_corkum1,centrifuge_corkum2,Mullin1,Mullin2,Valery},
``molecular propeller''~\cite{NJP,Kitano_udr,Khodorkovsky_udr_benzene},
and ``chiral train'' of laser pulses~\cite{Floss_prl}.

In this Letter, we investigate collisional dynamics in a  gas of molecules initially at room temperature,
but with extremely high rotational excitation. Our research is stimulated by recent laser experiments that achieved
the SR regime at near-ambient conditions~\cite{Mullin1,Mullin2,Valery,Bucksbaum}.
The theoretical analysis of these experiments is still in its infancy, and the previous quantum treatments focused on ultra-cold collisions of super-rotors~\cite{Forrey2001,Forrey2011}
are intractable in this case. Our study provides the first modeling (to the best of our knowledge)
of the collisional dynamics of molecular SRs by direct molecular dynamics (MD) simulation methods.
The classical approach seems to be an appropriate tool for studying molecules
at room temperature, excited to the quasiclassical rotational states with the angular momentum of tens to hundreds
of $\hbar$. It has already been successfully used for analysis of the laser experiments on molecular alignment at dissipative conditions~\cite{Hartmann12,Hartmann13b,Karras2014}.

We  studied relaxation dynamics for several small linear molecules (such $N_{2}$, $O_{2}$ and $CO_{2}$)
and found that they demonstrate a rather universal behavior in the
SR regime. Therefore, we  present below in detail our results for the representative
case of molecular nitrogen, $^{14}N_{2}$. The Supplementary Movie s1 (\url{https://www.youtube.com/watch?v=2qyoPvJ_SpI}) shows in an animated way a very similar relaxation dynamics in a gas of $O_2$ molecules.

\begin{figure}
\centering \includegraphics[width=0.5\textwidth]{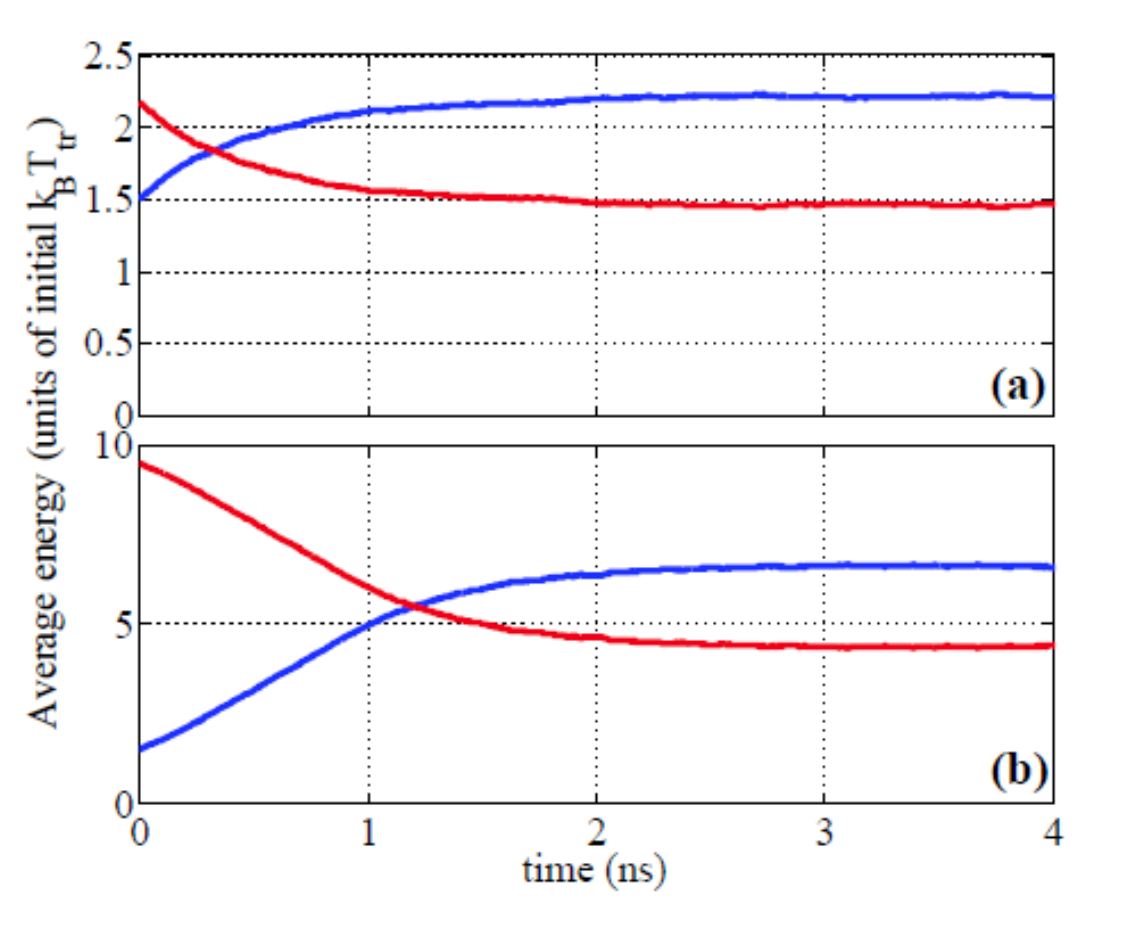}
\protect\caption{Equilibration of energy with time for a gas of $^{14}N_2$ molecules being initially at ambient temperature and pressure, and kicked by a linearly polarized laser pulse with a kick strength of (a) $P = 15$, and (b) $P=40$. The translational and rotational energy of the molecules are plotted in blue and red, respectively. }

\label{expo_and_nonexpo}
\end{figure}

We start by presenting results of a test case of a moderate rotational
excitation of molecules by a single non-resonant short laser pulse having linear polarization. Such a pulse is quantified by the so called
"kick strength" parameter, $P$ that defines a typical amount of the angular momentum (in units of $\hbar$  ) transferred from the laser pulse to a molecule (see the Methods section). The simulation of the initial rotational state by
laser excitation was performed as in Ref.~\cite{Khodorkovsky_udr_benzene}.
Figure \ref{expo_and_nonexpo}(a) demonstrates an exponential decay of the excess rotational energy of $N_2$ molecules after the pulse of $P=15$, and the corresponding increase of the translational energy as a function of time.
Such a kick strength (and even higher) can be achieved~\cite{Bucksbaum}  by a train of several laser pulses (50 fs duration, 800 nm central wavelength,  peak intensity of $36\ TW/cm^2$) separated in time by the rotational revival time $\sim 8.4$ ps of $^{14}N_2$ molecules. The magnitude of this kick  is comparable with the thermal value of the angular momentum quantum number  $J \sim 10$ for  nitrogen molecules at $T = 300$~K. At the end of the relaxation process, the ratio between the average translational and rotational energies is $3/2$, as required by the equipartition theorem.  The gas reaches equilibrium after about $2\,\text{ns}$ (after each molecule has experienced
about 20 collisions). Such an exponential relaxation pattern is in
agreement with other recent theoretical~\cite{Ramakrishna,Hartmann12}
and experimental~\cite{Hartmann13b,Karras2014,Owschimikow}
studies on the collisional relaxation of laser-excited molecular rotations.
For a stronger laser pulse of $P=40$, a slight deviation from the
exponential behavior appears at short times, see Fig.\ \ref{expo_and_nonexpo}(b).

\begin{figure}
\centering \includegraphics[width=0.5\textwidth]{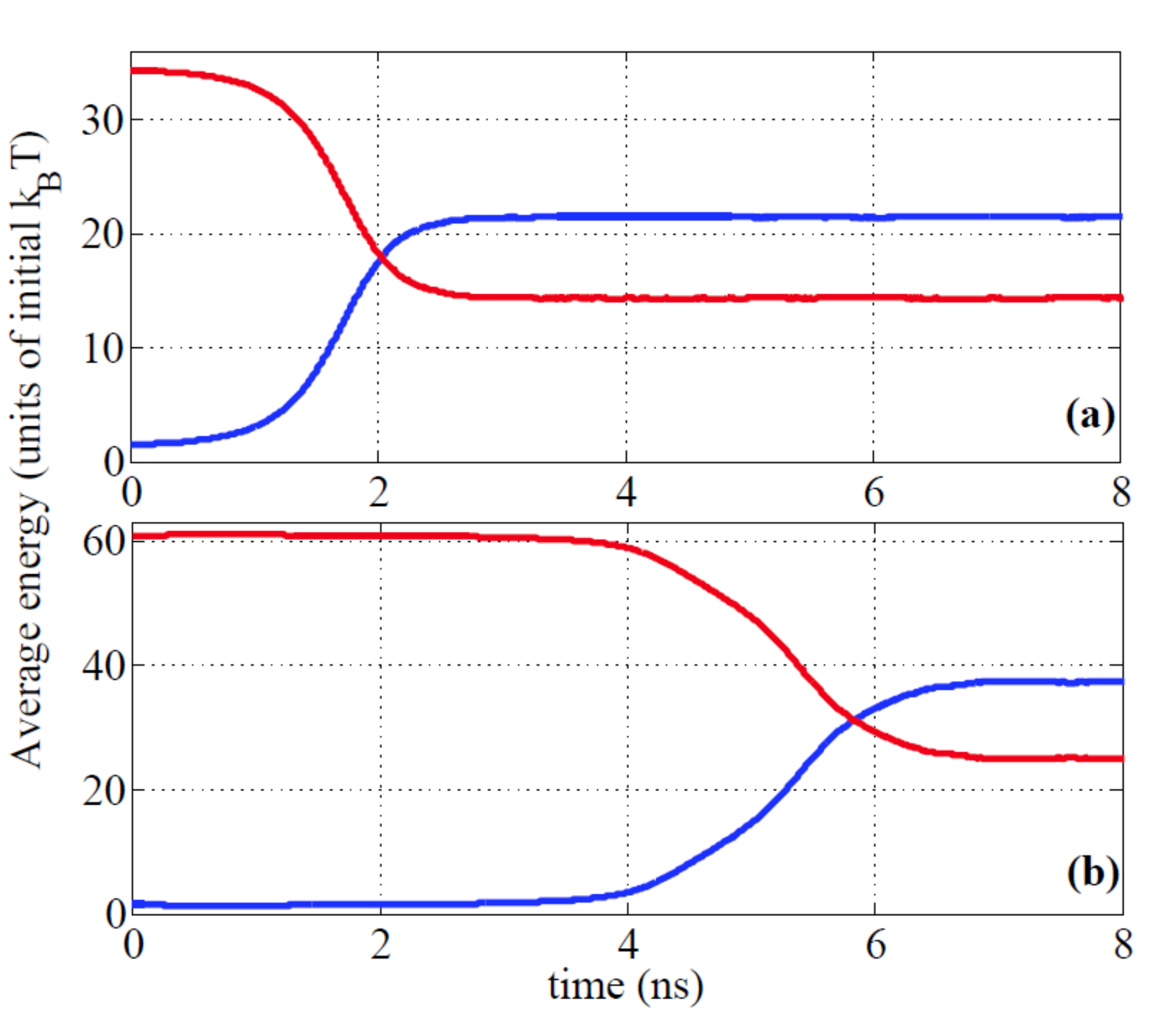}
\protect\caption{Equilibration of energy with time for a gas of $^{14}N_2$ molecules being  at ambient temperature and pressure, and subject to an optical centrifuge at $t=0$. Panels (a)and (b) correspond to the initial angular momentum of $J=60$ and $80$, respectively. The mean translational and rotational energies of the molecules are plotted in blue and red, respectively.}

\label{expo_and_nonexpo2}
\end{figure}

We now demonstrate that the relaxation scenario is drastically different for higher rotational excitation attainable by the ``optical centrifuge'' technique~\cite{centrifuge_corkum1,centrifuge_corkum2,Mullin1,Mullin2,Valery}.
Figure~\ref{expo_and_nonexpo2} presents the results of MD simulations
for an ensemble of nitrogen molecules that were initially aligned along the $x$-axis and
brought to fast rotation in the $xy$-plane with the same initial angular momentum  $\vec{J}$ oriented along the $z$-axis.
 This is a rather crude model for molecules suddenly released from an optical centrifuge, however simulations
using a more realistic description of the centrifugation process lead
to similar outcomes. Figure\ \ref{expo_and_nonexpo2} displays the
RT energy transfer dynamics for molecules initially excited to $J=60$
and $80$.

\begin{figure}
\centering

\includegraphics[width= 0.7\columnwidth]{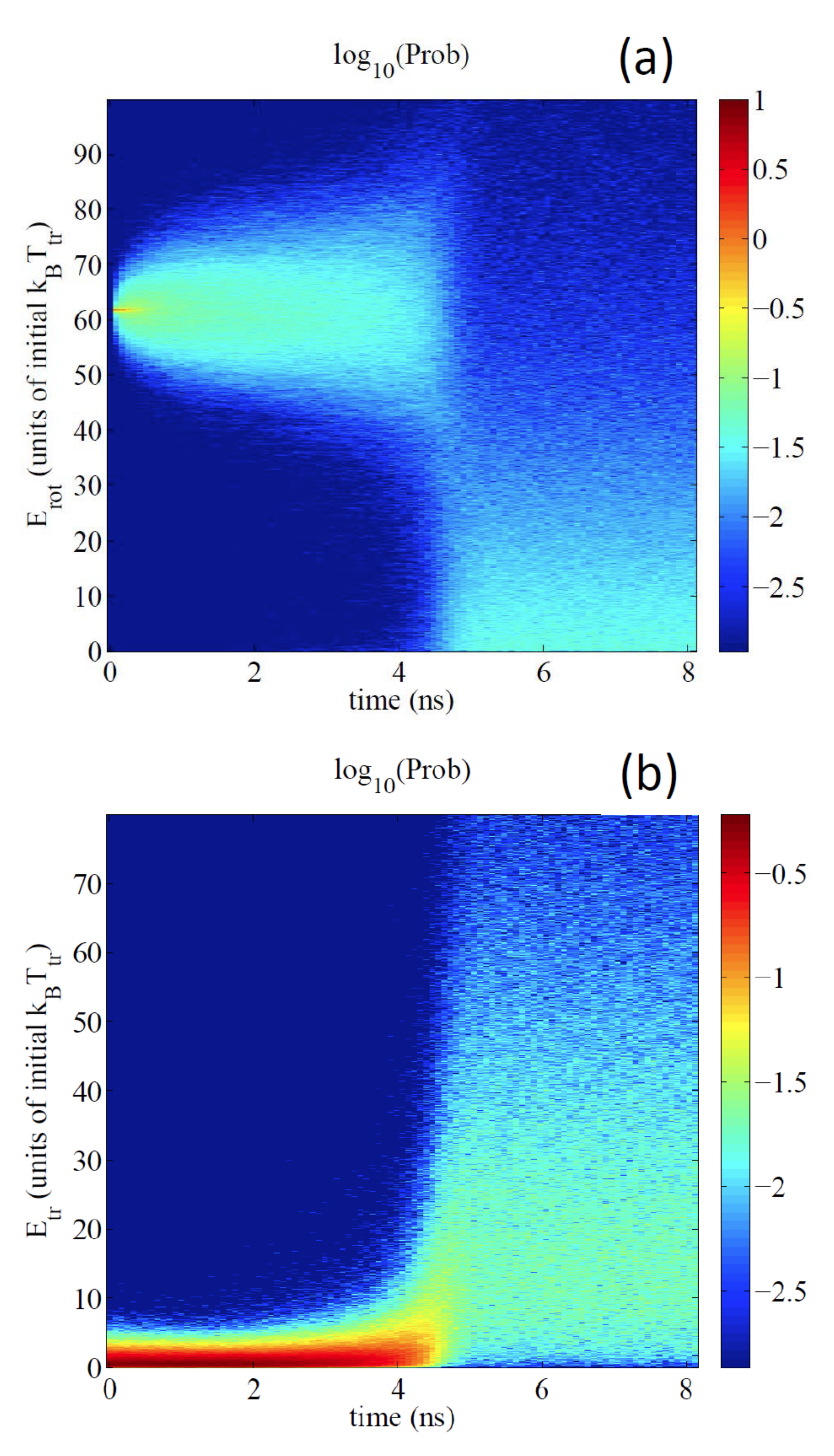}\protect\caption{Density plots for  time-dependent distributions of the rotational and translational energy (panels (a) and (b), respectively). The molecules are initially centrifuged to $J=80$. An explosive transition towards the thermal Maxwell distribution of the heated gas is clearly seen near 5 ns (compare with  Fig.\ \ref{expo_and_nonexpo2}(b)).
}

\label{energy_dist}
\end{figure}

Two distinct relaxation stages are clearly seen. Initially, for a relatively long period covering several tens of collisions,
there is almost no change in the mean rotational energy of the molecules (plateau region). After a certain induction period, the plateau is followed by an abrupt, explosive-like energy dump from the rotational motion into the translational one, leading to the final
thermal equilibration of the heated gas. The duration of the induction period grows with the initial rotational excitation. A deeper insight is given by the numerically simulated time-dependent distribution functions for the rotational (Fig.~\ref{energy_dist}(a)) and translational (Fig.~\ref{energy_dist}(b)) energies.  They show that during the plateau period, the rotational distribution broadens mainly due to the rotation-rotation energy exchange, and both distributions dramatically change their shapes in the course of the explosive relaxation. There is a clear physical reason for such relaxation dynamics. During the plateau period, the RT relaxation is inhibited due to the high value of the adiabaticity parameter, $a\gg1$. However, as the rotational distribution broadens, and the energy leaks from rotations to translations, there is an increasing fraction of molecules colliding at lower values of $a$. This makes the RT relaxation more efficient and triggers a positive feedback mechanism. The increase in the translational speeds shortens the typical collision time and raises the heating rate of the gas, which in turn accelerates further the  RT energy transfer and ignites an explosive thermalization. This mechanism resembles the famous phenomenon of thermal explosion~\cite{Semenov}  in heat-generating solids, which was first considered by N. Semenov in 1928. The Supplementary Movie s1 (\url{https://www.youtube.com/watch?v=2qyoPvJ_SpI}) presents in an animated way this relaxation scenario in the gas of  oxygen super-rotors. It shows the same generic relaxation stages (inhibited and explosive) as in the above example of the nitrogen gas. A shorter time scale is attributed to the higher gas pressure chosen for that numerical example.

\begin{figure}
\centering \includegraphics[width=0.5\textwidth]{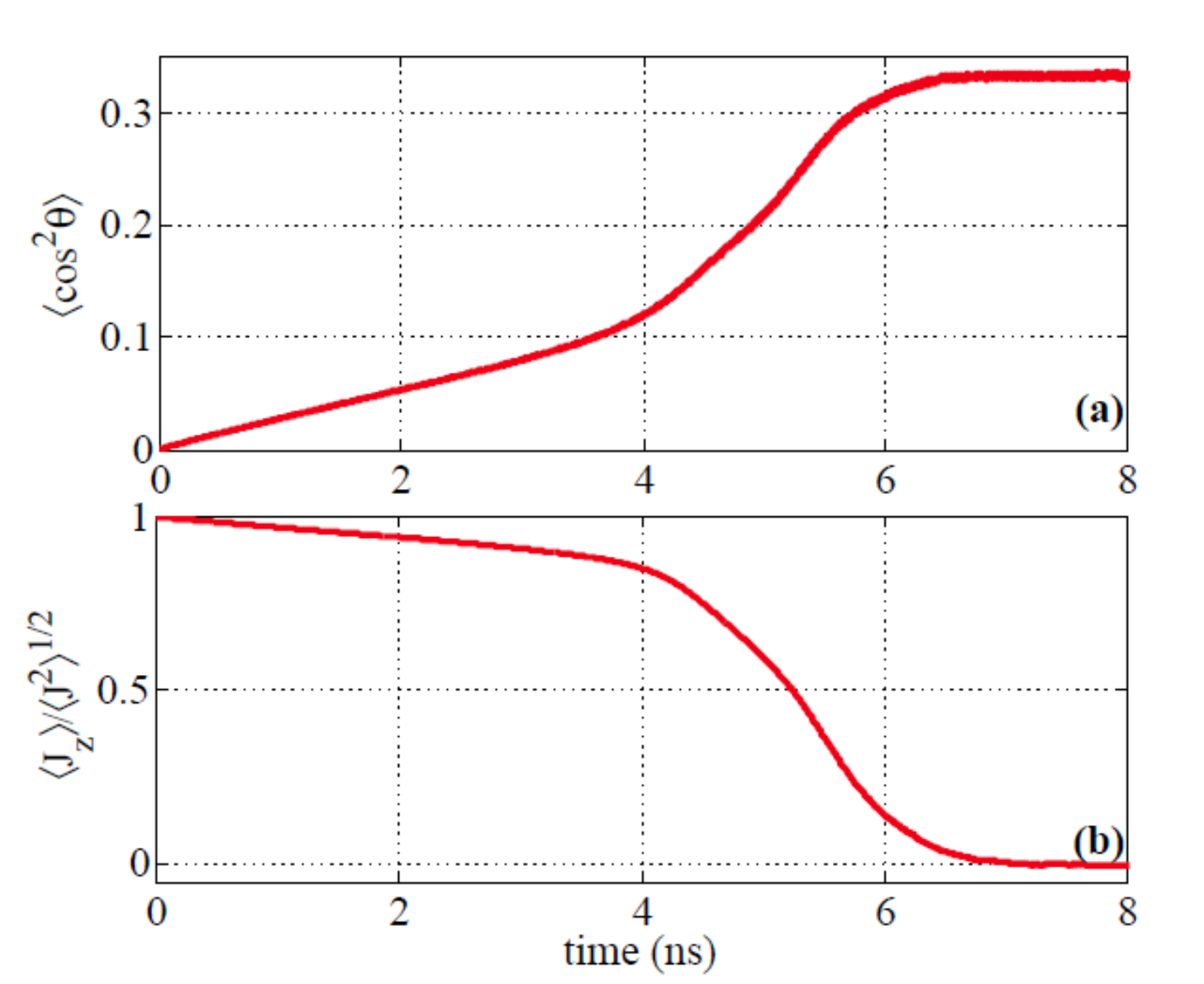}
\protect\caption{Orientation dynamics of the molecules that are initially centrifuged
in the $xy$-plane, $J=80$. Panel (a)- time dependence of the alignment
factor along $z$ axis. Panel (b) - decay of the normalized $z$
component of the angular momentum.}

\label{orient}
\end{figure}

To get a deeper insight into the collisional relaxation of SRs, we investigated
the process of isotropization of the  molecular orientation, and of the direction of
the rotational angular momentum. Figure \ref{orient}(a) shows that
during the plateau period (until about $4\,\text{ns}$), the factor
of alignment along the $z$-axis, $\left\langle \cos^{2}\theta\right\rangle $,
remains close to zero. This means that the molecules continue rotating
mainly in the $xy$-plane with the angular momentum vector oriented
along the $z$-axis. This is also supported by Fig.\ \ref{orient}(b)
that shows the time dependence of the $z$ component (parallel to
the exciting laser's propagation direction) of the molecular angular
momentum, $J_{z}$. The ability to keep orientation of the angular
momentum is characteristic of gyroscopes (even as small as a rotating
molecule). Such a tendency is consistent with a much weaker propensity observed in recent studies
of laser excited molecules \cite{Hartmann12} at lower degrees of rotational excitation.
For this reason, we term the whole period before the self-accelerating
RT energy exchange, as a ``gyroscopic stage'' of relaxation. As
seen from Fig.\ \ref{orient}, the isotropization of the super-rotors
occurs mainly during the stage of the explosive relaxation, after
which the alignment factor takes the isotropic value of $\left\langle \cos^{2}\theta\right\rangle =1/3$,
and there is no preferred orientation of the angular momentum vector.

\begin{figure}
\centering

\includegraphics[width=0.5\columnwidth]{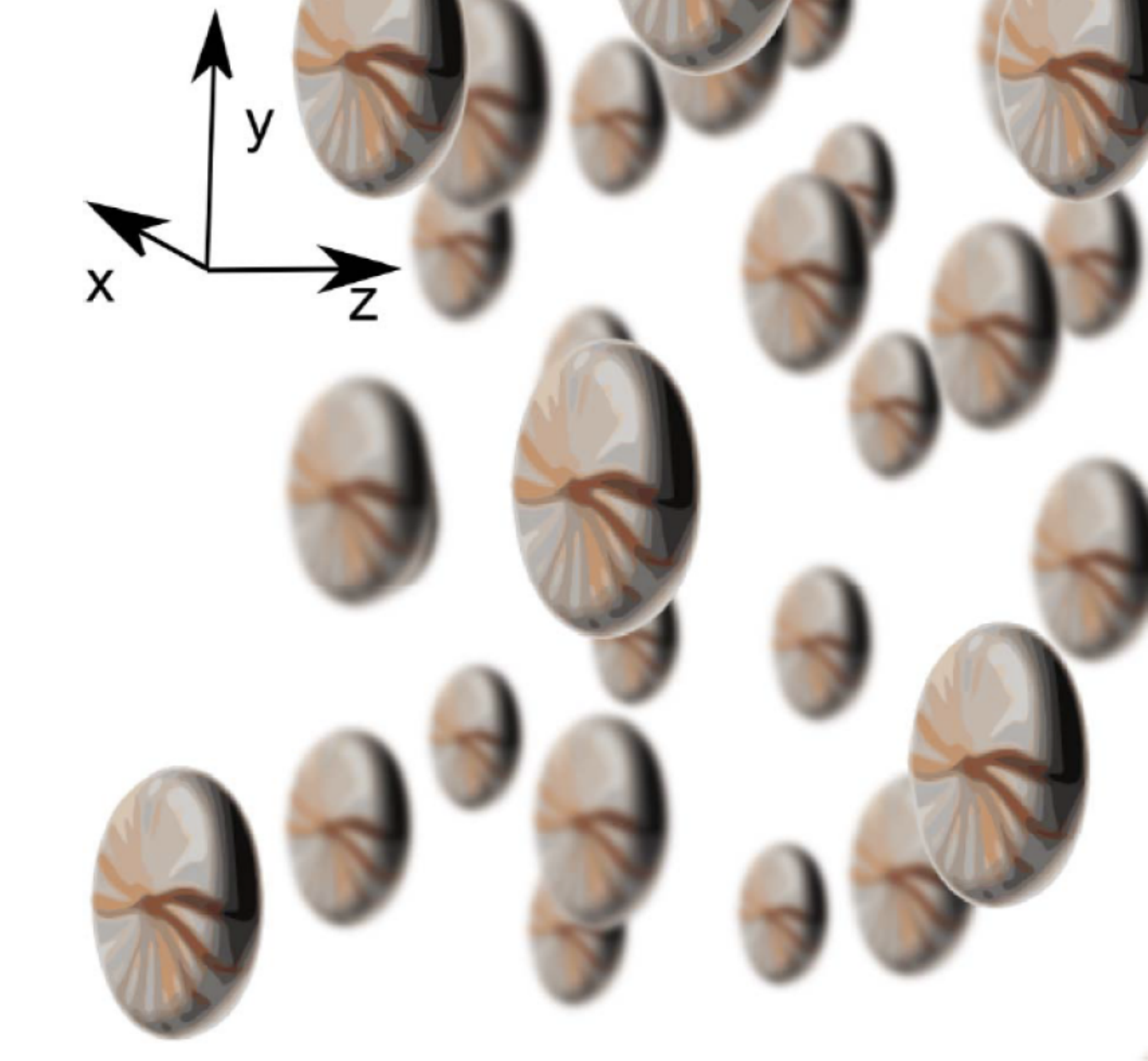}\protect\caption{Schematic illustration of the anisotropic angular distribution of gas molecules spun by an optical centrifuge rotating in the $xy$-plane. }

\label{fig_pancakes}
\end{figure}

Figure~\ref{fig_pancakes} schematically shows the anisotropic (pancake-like) angular distribution maintained by the molecular super-rotors in the course of many collisions during the gyroscopic stage. The initial fast rotation was set by an optical centrifuge operating in the $xy$-plane.
The macroscopic polarizability of such a gas is anisotropic as well, which can be detected by measuring  birefringence for a polarized light propagating along any direction in the $xy$-plane. Such a persistent  birefringence may serve as an optical indicator of the gyroscopic relaxation stage.  However, here we will focus on the gas-kinetics manifestations of this anisotropy. Just looking at Fig.\ \ref{fig_pancakes},
one may expect that the collisional cross-section for the face-to-face colliding molecular `pancakes' should be different from the  cross-section
for the side-to-side collisions. Also, the rotational-translational energy exchange happens differently in the cases of the face-to-face and side-to-side collisions~\cite{Mullin2}.  This suggests the emergence of  anisotropic transport phenomena in the gas, such as  anisotropic mass and energy diffusion.  Moreover, the explosive RT energy exchange at the end of the gyroscopic phase should have its manifestations
in the transport effects as well.

\begin{figure}
\centering \includegraphics[width=\columnwidth]{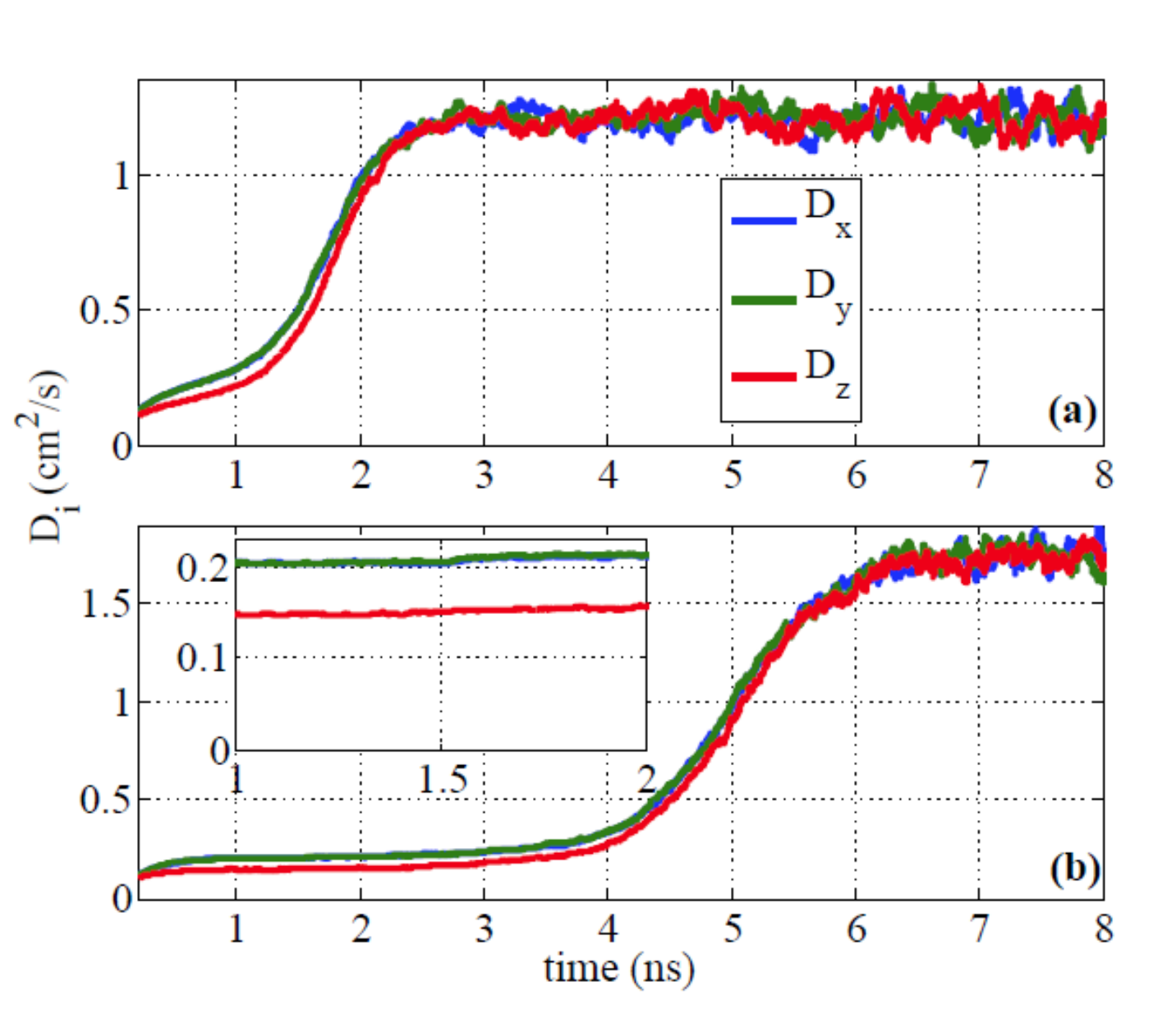}\protect\caption{Time-dependent diffusion coefficients $D_{i}(t)$ along the axes $i=x,y,z$ are plotted as a function of time. The molecules in panels (a) and (b) are initially centrifuged in the $xy$-plane with an angular momentum of $J=60$ and $J=80$, respectively.}

\label{diffusion}
\end{figure}

To test these predictions, we investigated the process of self-diffusion of a  nitrogen molecular super-rotor by means of the above direct MD simulations. Following the random-walk trajectory for each molecule of an ensemble prepared in a centrifuged state, we calculated the average square of the cumulative molecular displacement along each of the Cartesian axes as a function of time. The time-dependent  diffusion coefficients for the motion along every axis was derived by differentiating this function with respect to time (e.g.,  $D_x(t)=\frac{1}{2} \frac{d\langle (\Delta x)^2\rangle}{dt}$).
Figure~\ref{diffusion}(b) shows that as long as the molecules keep the original orientation of their rotation axis (gyroscopic stage), the diffusion coefficients in the $x$ and $y$ directions are indeed about $40\% $ larger than the one corresponding to the diffusion along the $z$ axis. The diffusion coefficients grow with time (as seen
in Fig.\ \ref{diffusion}) because of the gradual heating of the translational motion of the molecules
due to the RT energy exchange. With the ignition of the explosive relaxation,
the diffusion coefficients grow abruptly together with the translational speeds, and all three coefficients
reach the same high value corresponding to the final hot equilibrated
state.

Nontrivial regimes of mass and heat transfer in the gas of
super-rotors may happen when the laser excitation takes place in a
spatially-confined domain, like the focal spot of a focused laser
beam, which implies that the `cloud' of centrifuged molecules disperses
differently in the axial and radial directions. The explosive-like delivery of the
rotational energy to the translational motion leads to an increased local
pressure and to the subsequent generation of sound waves, as well as to the creation of a hot
area with a reduced molecular concentration which may be optically
observed \cite{Steinitz,Zahedpour2014,Segev}. This density hole is then
slowly healed via the diffusion of cold unexcited molecules from the
area outside the focal spot, while the hot molecules diffuse outwards. The multi-scale long decay of rotational
signals from the optically centrifuged molecules in dense gases reported
recently \cite{Mullin1,Mullin2} may be actually related to the above interplay of the
kinetic and hydrodynamic phenomena, which requires additional theoretical
analysis of these experiments.

Finally, we emphasize that although the molecular gas arrives to a
microscopically thermalized state, macroscopic non-equilibrium
flows can still exist in the gas. For molecules that are initially excited
into a rotation with a preferred sense of spinning (with oriented  mean angular momentum), it was shown recently that a vortex
flow appears around the laser focal spot  after thermalization~\cite{Steinitz}.

To summarize, we investigated theoretically the collisional relaxation in a gas of
molecular super-rotors, excited  with the help of ultrashort
laser pulses. We found that the SRs behave like tiny gyroscopes  after the excitation, and  maintain their fast rotation and orientation of the angular momentum through many collisions. This leads to the anisotropic diffusion in the gas, and to a long-lasting optical birefringence.
The rotational-translational energy transfer is effectively suppressed during the gyroscopic stage. This metastable state of the gas terminates abruptly after some induction period (depending on the initial rotational excitation) due to development of an explosive self-accelerating energy transfer from the rotational degrees of freedom to the translational ones. We discussed optical means for detecting the macroscopic effects associated with SRs relaxation. Other manifestations may include filamentation effects, anisotropic diffusion of selected species in a molecular mixture of different components, anisotropic viscosity in the gas, and directed acousto-optical signals produced by the laser-generated rotationally excited gas \cite{Schippers}.

\section{Methods}
The simulations were carried
out with $\sim10^{4}$ molecules residing in a cubic volume at room
translational temperature and pressure. Initially, the molecules were uniformly
distributed in space, while their initial velocity
followed the Maxwell-Boltzmann distribution. The  initial distribution of the rotational
degrees of freedom (direction of molecular axes, angular velocity
vectors) was obtained by simulating the process of ultra-fast laser
excitation of SRs. In the case of the excitation by a linearly polarized aligning laser pulse (see Fig.\ \ref{expo_and_nonexpo}), the details of the simulation procedure are described in Ref.~\cite{Khodorkovsky_udr_benzene}. Specifically, the molecules were impulsively excited as if affected by a linearly polarized femtosecond laser pulse with a kick strength $P=\Delta\alpha\int_{-\infty}^{\infty} \mathcal{E}^2(t)\,dt/(4\hbar)$ (where $\Delta\alpha$ is the molecular polarizability anisotropy and $\mathcal{E}$ is the envelope of the electric field of the laser pulse). Parameter $P$ defines a typical amount of the angular momentum (in units of $\hbar$  ) transferred from the laser pulse to a molecule. Excitation by the optical centrifuge was simulated by solving classical equations of motion for a thermal ensemble of  polarizable linear molecules subject to a pulsed electric field with rotating polarization.  The parameters of the centrifuge were similar to those used in the experiment \cite{Valery}.  Starting from these initial conditions, we let
the molecules propagate in space using the standard Verlet algorithm,
and periodic boundary conditions, while the propagation of the rotational
motion was done as described in Ref.~\cite{Allen}. The classical intermolecular
interaction potential was taken from Ref.~\cite{Cheung}.

\section{Acknowledgements}
 This research was made possible in part by the historic generosity of the Harold Perlman Family.
 I.A. acknowledges support as the Patricia Elman Bildner Professorial Chair. He also acknowledges kind hospitality and support from the Universit\'{e} Paris Est Cr\'{e}teil during a one week stay with J-M.H. at LISA.

\section{Author contributions}
All authors contributed to the formulation of the problem and analysis of the results. Numerical simulations were mainly performed by Y.K. and U.S.. I.A. and J-M.H. supervised the research.  All authors discussed the results and the manuscript.

\section{Additional information}
{\bf Supplementary Movie s1.}

The movie can be found on \url{https://www.youtube.com/watch?v=2qyoPvJ_SpI}

\underline{Animated visualization of the collisional dynamics in a gas of oxygen super-rotors.}  This video depicts the simulation of $\sim 200$ spinning oxygen molecules, initially at room translational temperature, all rotating fast at $J\sim 60$. For visualization purposes the initial concentration is 20 times that of atmospheric gas and the boundary conditions are periodic. The resulting dynamics follows the same generic scenario as observed in the molecular dynamics simulations for nitrogen molecules in the main text of the paper, only faster by a factor of about 20 (due to the increased pressure). In the left box each particle is represented by a spindle whose axis corresponds to the rotation axis, and its length depicts the rotation speed. The color of each particle codes the translational speed. On the right panel, the evolution of the mean translational and rotational energy  is shown (compare with Fig.\ \ref{expo_and_nonexpo2} in the main text), along with the momentary distribution of the rotational energy (green markers). The first quarter of the movie features the metastable gyroscopic stage, where the molecules' rotation axes are still parallel to the original preferred direction, despite the many collisions each of them has undergone. The stage is followed by the explosive relaxation (at around halfway through the movie) where the molecules suddenly transform from slow (blue) parallel spindles into fast moving (red) ones with slowed, isotropic rotation.

\section{Competing financial interests}
The authors declare no competing financial interests.

\newpage

%\clearpage
%\clearpage
%\clearpage
%\clearpage
%\clearpage

\end{document}